\documentclass[nofootinbib,amsmath,amssymb,showpacs,showkeys,aps,reprint,prd]{revtex4-1}
\usepackage[utf8]{inputenc} 
\usepackage{amssymb}
\usepackage{graphicx}
\usepackage{dcolumn}
\usepackage{subfig}
\usepackage[dvipsnames]{xcolor}
\usepackage[T1]{fontenc}

\usepackage{mathrsfs}  
\usepackage{cases}
\usepackage{physics}
\usepackage{bm}
\usepackage{academicons}
\usepackage{mathtools, nccmath}
\usepackage{fancyhdr}
\usepackage{tikz,xcolor}

\usepackage{tensor}
\usepackage[normalem]{ulem}
\usepackage{lipsum}
\usepackage{soul}
\usepackage{cancel}
\usepackage{stackengine,scalerel}
\usepackage{hyperref}
\usepackage{tabularx}
\usepackage[justification=raggedright]{caption}
\usepackage{placeins}
\hypersetup{colorlinks, linkcolor={red},citecolor={blue},urlcolor={blue}}  

\usepackage{orcidlink}

\def\di{\mathrm{d}}
\def\bol{\mathrm{bol}}
\def\eff{\mathrm{eff}}
\def\isco{\mathrm{ISCO}}

\newcommand{\be}{\begin{equation}}
\newcommand{\ee}{\end{equation}}

\newcommand{\mincir}{\raise
-3.truept\hbox{\rlap{\hbox{$\sim$}}\raise4.truept\hbox{$<$}\ }}
\newcommand{\magcir}{\raise
-3.truept\hbox{\rlap{\hbox{$\sim$}}\raise4.truept\hbox{$>$}\ }}

\providecommand{\U}[1]{\protect\rule{.1in}{.1in}}

\begin{document}
\title{\Large Black holes surrounded by massive vector fields in Kaluza-Klein gravity }
\author{Kimet Jusufi\orcidlink{0000-0003-0527-4177}$^1$}
\email[]{kimet.jusufi@unite.edu.mk (corresponding author)} 
\author{Ankit Anand\orcidlink{0000-0002-8832-3212}$^2$}
\email[]{anand@iitk.ac.in} 
\author{Sara Saghafi\orcidlink{0000-0001-8381-973X}$^4$} 
\email[]{s.saghafi@umz.ac.ir}
\author{B. Cuadros-Melgar\orcidlink{0000-0001-8111-2431
}$^3$}
\email[]{bertha@usp.br}
\author{Kourosh Nozari\orcidlink{0000-0003-4368-5823}$^4$}
\email[]{knozari@umz.ac.ir}
\affiliation{$^1$Physics Department, University of Tetova, Ilinden Street nn, 1200, Tetovo, North Macedonia}
\affiliation{$^2$Department of Physics, Indian Institute of Technology, Kanpur 208016, India}
\affiliation{$^{3}$Engineering School of Lorena - University of Sao Paulo (EEL-USP), Estrada Municipal do Campinho N$^o$ 100, Campinho, CEP: 12602-810, Lorena, SP - Brazil}
\affiliation{$^{4}$Department of Theoretical Physics, Faculty of Science, University of Mazandaran, P. O. Box 47416-95447, Babolsar, Iran}

%%%%%%%%%%%%%%%%%%%%%%%%%%%%%%%%%%%%%%%%%%%%%%%%%%%%%%%%%%%%%%%%%%
\begin{abstract}
We present an exact black hole solution surrounded by massive vector fields predicted by Kaluza-Klein (KK) gravity. KK gravity in four dimensions (4D) is of particular interest, as it predicts a tower of particle states, including gravitons with spin-0 and spin-1 components, in addition to the massless spin-2 gravitons of general relativity. The extra degrees of freedom in the gravitational sector modify the law of gravity, allowing the theory to explain the effects attributed to dark matter in the universe.  In this paper, we construct a black hole solution surrounded by massive spin-1 gravitons within KK theory. In addition to the influence of the massive vector fields, we incorporate an interaction term between the black hole and the massive vector field. The black hole solution is affected by the mass of the spin-1 graviton and an additional parameter that encodes corrections to Newton's constant, as well as the coupling between the massive vector field and the black hole mass. We find that the coupling between the massive vector field and the black hole mimics the effect of an electric charge. To this end, we investigate the accretion disk, quasinormal modes (QNMs), and the stability of the black hole spacetime. Finally, we use Event Horizon Telescope (EHT) observations of Sgr A* to constrain the black hole parameters.

\end{abstract}

\maketitle
%\tableofcontents

%%%%%%%%%%%%%%%%%%%%%%%%%%%%%%%%%%%%%%%%%%%%%%%%%%%%%%%%%%%%%%%%%%%%%%%%%%%%%%%%%%%%%%%%%%%%%%%%%%%5
\section{Introduction}

Black holes (BHs) are extraordinary astrophysical objects that exhibit the effects of extreme gravity and high-energy physics. These effects manifest in phenomena such as the formation of colossal particle jets, quasi-periodic oscillations, gravitational lensing, and the disruption of nearby stars.  Over the past few decades, groundbreaking observations have confirmed BHs compellingly. From X-ray emissions \cite{Fabian:1989ej}, radio signals \cite{Alexander:2005jz} from regions near BHs, the imaging of the event horizon shadow of the M87 supermassive BH by the EHT Collaboration (2019), and the detection of gravitational waves from BH mergers by the LIGO/Virgo Collaboration \cite{LIGOScientific:2016wkq} have all contributed to this growing body of evidence. Theoretically, BHs serve as critical testing grounds for exploring the predictions of modified gravity, quantum gravity, and other corrections to general relativity, offering invaluable insights into the fundamental nature of spacetime and the universe. On cosmic scales, photons originating from a luminous source near a black hole (BH) follow trajectories that either lead to capture by the event horizon or escape to infinity. The boundary between these two behaviors is determined by unstable spherical photon orbits, which define the black hole's shadow \cite{Takahashi:2004xh, Hioki:2009na, Brito:2015oca, Cunha:2015yba, Ohgami:2015nra, Moffat:2015kva, Abdujabbarov:2016hnw, Cunha:2018acu, Mizuno:2018lxz, Tsukamoto:2017fxq, Psaltis:2018xkc, Chen:2022nbb,Amir:2018pcu, Gralla:2019xty, Bambi:2019tjh, Cunha:2019ikd, Khodadi:2020jij, Khodadi:2024ubi,Perlick:2021aok, Vagnozzi:2022moj, Saurabh:2020zqg, Jusufi:2020cpn, Tsupko:2019pzg}. The observational reconstruction of such trajectories, as achieved by the EHT Collaboration, provides direct insight into the nature of black holes. In addition, a fundamental connection between black hole shadows and quasinormal modes (QNMs) in the eikonal regime has been theoretically proposed and analytically confirmed~\cite{Jusufi:2019ltj, Cuadros-Melgar:2020kqn,Pedrotti:2024znu}. 

At the same time, the Yukawa potential has emerged as a significant modification to gravitational interactions, offering a viable extension to the standard \(\Lambda\)CDM framework in cosmology \cite{Jusufi:2023xoa, Gonzalez:2023rsd}. This formulation establishes a self-consistent relation between baryonic matter, effective dark matter, and dark energy, suggesting that deviations from Newtonian gravity may give rise to the observed dark matter effects rather than requiring it as a fundamental entity. Specifically, in the absence of baryonic matter, the effective dark matter component vanishes, reinforcing the hypothesis that dark matter manifests as an emergent phenomenon induced by gravitational corrections. Empirical analyses confirm that the Yukawa-modified cosmology can reproduce observationally consistent values for the dark energy and dark matter density fractions \cite{Gonzalez:2023rsd}. Extending this framework to black hole physics \cite{Filho:2023abd,FILHO2024101711}, recent studies have incorporated Yukawa corrections into the gravitational and electromagnetic potentials governing charged black holes, revealing nontrivial modifications to their geometric, thermodynamic, and stability properties. In addition, the accretion process is the gravitationally driven inflow of matter onto a black hole, facilitating the redistribution of angular momentum and the conversion of gravitational potential energy into radiation \cite{Shakura:1972te,novikov1973astrophysics,page1974disk,thorne1974disk}.

Accretion is the process by which matter gravitationally falls onto a compact object, such as a black hole, redistributing energy and driving high-energy astrophysical phenomena like relativistic jets and quasars \cite{2008bhad.book}. Accretion disks are geometrically thin, optically thick structures, which form as infalling gas conserves angular momentum, initially following metastable orbits before transitioning into unstable trajectories dictated by spacetime geometry. Gravitational energy converts into thermal radiation through viscous dissipation and magnetohydrodynamic interactions, with emissions peaking in the inner disk regions where relativistic effects dominate. The radiation spectrum, spanning from radio to X-ray wavelengths, depends on the velocity distribution of accreting matter and the curvature of spacetime. Key geodesic structures, including photon orbits ($r_{ph}$) and the innermost stable circular orbit (ISCO), define the limits of stable motion. In thin disks, the ISCO marks the inner edge beyond which matter plunges into the event horizon, constraining black hole properties and accretion efficiency. Perturbations from pressure variations, magnetic fields, or relativistic frame-dragging induce epicyclic oscillations with radial and vertical components. These oscillations in the inner disk contribute to quasi-periodic variability observed in black hole systems and influence emitted spectra. Understanding orbital motion and epicyclic frequencies is crucial for probing relativistic accretion dynamics and black hole spacetime geometry. Black hole accretion disks \cite{Jiao:2016iwp, Johannsen:2013asa, John:2013bqa, Karkowski:2012vt, Liu:2020vkh, Mach:2013gia, Murtaza:2024ylz, Mustafa:2023scj, Nozari:2020swx, Nozari:2023enj, Ortega-Rodriguez:2006kdi, Yuan:2014gma, Zheng:2019mem, Nozari:2024vxp} have been a subject of extensive research and have garnered significant attention in the scientific literature.

On the other hand, perturbations in black hole spacetimes give rise to gravitational waves characterized by complex eigenfrequencies, termed  QNMs. These modes encode fundamental information about the dynamical response of the black hole to external disturbances. The real part of the QNMs frequency determines the characteristic oscillation frequency of the emitted gravitational radiation. In contrast, the imaginary part dictates its exponential decay or amplification, governing the stability of the perturbations. The foundation of black hole perturbation theory was established by studying axial perturbations in the Schwarzschild metric using a linearized approach \cite{Regge:1957td}. This was later extended to polar perturbations, ensuring the energy-momentum tensor remained unaffected \cite{Zerilli:1974ai}. The significance of QNMs in black hole dynamics was later recognized, leading to a systematic mathematical treatment of their spectral properties \cite{Vishveshwara:1970cc, Vishveshwara:1970zz, Chandrasekhar:1975zza}. Subsequent research expanded QNMs analyses across various gravitational frameworks, including general relativity, alternative gravity theories \cite{Konoplya:2004xx, Natario:2004jd, Nagar:2005ea, Nagar:2005ea, Konoplya:2006rv, Abdalla_2006, Cuadros_Melgar_2012, Abdalla_2019, Ferrari:2000ep, Li:2001ct}, and nonsingular black hole solutions \cite{Toshmatov:2015wga, Panotopoulos:2019qjk}. These investigations have provided critical insights into black hole stability, ringdown behavior, and potential observational signatures in astrophysical and high-energy contexts. A potential link exists between the thermodynamic behavior of loop quantum black holes and the quasinormal oscillations observed in astrophysical black hole candidates \cite{Konoplya2003, Konoplya:2011qq}. Moreover, examining QNMs in the context of gravitational collapse could yield significant insights \cite{Purrer:2004nq}. The interplay between black hole QNMs and shadows has gained interest, especially after the first black hole image by the EHT \cite{EventHorizonTelescope:2019dse, EventHorizonTelescope:2019pgp, EventHorizonTelescope:2019ggy}. Extensive studies on shadow properties \cite{Amir:2018szm, Amir:2018pcu, Jusufi:2019nrn, Jusufi:2020cpn, Haroon:2019new, Vagnozzi:2020quf} and gravitational wave detections \cite{LIGOScientific:2016aoc} further motivate this connection, offering new insights into black hole physics. Various analytical and numerical techniques have been developed to study them, with the Wentzel-Kramers-Brillouin (WKB) approximation being one of the most widely utilized methods \cite{Schutz:1985km, Iyer:1986np, Konoplya2003, Matyjasek:2017psv, Toshmatov:2015wga}. The WKB approximation method determines QNMs by matching asymptotic solutions with a Taylor expansion around the peak of the effective potential.

In this paper, we aim to find a new black hole solution that is surrounded by massive vector fields in the KK theory, which has recently been proposed as an alternative to the dark sector \cite{Jusufi:2024utf}. This model introduces a generalized gravitational potential combining Yukawa and Newtonian terms and brings in additional degrees of freedom, namely spin-0 and spin-1 gravitons, which represent  the gravitational analogs of  gauge fields \cite{Jusufi:2024utf,Jusufi:2025hte}. We mention here that self-interacting massive vector fields often face instabilities and apparent pathologies \cite{Rubio:2024ryv}. This work also pursues to systematically investigate the influence of the massive vector field on the accretion disk, the quasinormal mode spectrum, and the photon sphere.

We briefly review the KK theory in Section \ref{Sec:Yukawa potential}. In Section \ref{Sec:Black hole solution surrounded by massive vector fields} we find the black hole solution. In Section \ref{Sec:Accretion disk} we explore the accretion disk around such black holes. In Section \ref{Sec:Quasinormal Modes and Stability} we explore the QNMs and the dynamical stability. In Section \ref{Sec:QNMsand shadow} we elaborate on the photon sphere and the shadow radius. Finally, in Sec.\ref{Sec:Conclusions} we conclude and discuss our findings.
%%%%%%%%%%%%%%%%%%%%%%%%%%%%%%%%%%%%%%%%%%%%%%%%%%%%%%%%%%%%%%%%%%%%%%%%%%%%%%%%%%%%%%%%%%%%%%%%%%%%%%%%%%%%%%%%%%%%%%%%%%%%%%%%%%%%%%%%

\section{Review of Kaluza-Klein gravity}\label{Sec:Yukawa potential}

In this section we review KK gravity based on the framework from Ref.~\cite{Jusufi:2024utf}, focusing on an alternative interpretation of KK theory that excludes electromagnetism. We consider a simplified KK model in $D=5$ dimensions with a generalized Einstein-Hilbert action and matter fields,
\begin{equation}
S_{\text{5D}} = \frac{1}{16 \pi \tilde{G}}\int d^4x \,dy
\sqrt{-\tilde{g}_{AB}} \tilde{R}+S_{\rm matter} \ ,
\end{equation}
Here, $y$ denotes the compactified dimension, while $\tilde{G}$, $ \tilde{R} = \tilde{g}_{AB} \tilde{R}^{AB}$, and $\tilde{g}_{AB}$ represent the five-dimensional gravitational constant, Ricci scalar, and metric tensor, respectively. The matter contribution $ S_{\rm matter}$ is expressed in terms of the four-dimensional spacetime metric $g_{\mu \nu}$, the scalar field $\Phi$, and the gauge field $A_{\mu}$,
\begin{equation}
  \tilde{g}_{AB} =\left( \begin{matrix}
g_{\mu\nu} + \Phi^2 A_\mu A_\nu & \Phi^2 A_\mu \\[2mm]
\Phi^2 A_\nu & \Phi^2
\end{matrix}\right) \ .
\end{equation}

After performing the dimensional reduction from $D=5$, we obtain a theory similar to a scalar-tensor model with an additional gauge field. The matter contribution $S_{\rm matter}$ can include a complex scalar field $\tilde{\phi}$ with a self-interaction potential $V(\tilde{\phi})$. Assuming that $\Phi$ is a slowly varying field, we can write the perturbation as $\Phi = \Phi_0 + \delta \Phi$, leading to $G \equiv G_N(1+\alpha)$, where $\alpha$ is a parameter. A key feature of KK gravity is its analogy to superconductivity, where photons acquire mass through spontaneous symmetry breaking \cite{Jusufi:2024utf}. The scalar field $\tilde{\phi}$, minimally coupled to the gauge field $A_\mu$ with coupling constant $g$, is rewritten as $\tilde{\phi} = \tilde{\phi}_0 e^{i\chi} $, where $\tilde{\phi}_0$ is the vacuum expectation value. Applying the Anderson-Higgs mechanism, the gauge-invariant Lagrangian in the KK model is given by,
\begin{equation}
\mathcal{L} =\frac{1}{16 \pi G} R-\dfrac{1}{4} \tilde{F}^{\mu \nu }\tilde{F}_{\mu
\nu }  - \frac{1}{2}\mu^2 \tilde{A}^{\mu }\tilde{A}_{\mu }-V(\tilde{\phi}) \ .
\label{KK-Lagr}
\end{equation}%
The gauge boson mass is given by $\mu^2 \equiv g^2 |\tilde{\phi}_0|^2$, where $V(\tilde{\phi})$ is the scalar field's self-interaction potential and $\tilde{A}_\mu$ is related to $A_\mu$ via the gauge transformation $\tilde{A}_\mu = A_\mu - \frac{\partial_\mu \chi}{g}$. The \( U(1) \) symmetry breaking in superconductors generates the mass term \( \mu^2 \tilde{A}_\mu \tilde{A}^\mu / 2 \), with the degree of freedom of the scalar field \( \tilde{\phi} \) being absorbed by the gauge boson. Consequently, there are two degrees of freedom for the massless spin-2 graviton and three for the massive spin-1 graviton. The variations of the gravitational Lagrangian with respect to the metric yield the Einstein field equations \cite{Jusufi:2024utf},
\begin{equation}
G_{\mu \nu}+\Lambda g_{\mu \nu}= 8 \pi G \left( T_{\mu \nu}^{\rm V}+T_{\mu \nu}^{\rm M}\right) \ ,
\end{equation}
with $T_{\mu \nu}^{\rm M}$ being the matter part of the stress-energy tensor, and $T_{\mu \nu}^{\rm V} $ the energy-momentum tensor for massive spin-1 graviton,
\begin{eqnarray}\notag\label{emtensor}
T_{\mu \nu}^{\rm V} &=& \frac{1}{4 \pi} \Big[\tilde{F}_{\mu \sigma} {\tilde{F}_\nu}^\sigma-\frac{1}{4} g_{\mu \nu} \tilde{F}^2\\
&+&\mu^2 \left(\tilde{A}_\mu \tilde{A}_\nu - \frac{1}{2}g_{\mu \nu} \tilde{A}_\sigma \tilde{A}^\sigma \right)\Big] \ .
\end{eqnarray}
In the strong-gravity regime we expect a contribution from the coupling between the vector field and the spacetime geometry, leading to a correction term (or interaction term) in the energy-momentum tensor, $T^{\rm int}_{\mu \nu}(g_{\mu \nu}, A_{\mu})$, which will be important in the next section. From the Lagrangian \eqref{KK-Lagr}, the motion of the vector field is governed by the Proca equation as \cite{Jusufi:2024utf},
\begin{equation}
\nabla_\mu \tilde{F}^{\mu\nu}-\mu^2 \tilde{A}^\mu = 0 \ ,
\end{equation}%
which by choosing the Lorenz gauge, $ \nabla_{\mu}\tilde{A}^{\mu}=0$, and after some calculations, leads to a wave equation for $\tilde{A}^{\mu}$, 
\begin{eqnarray}\label{waveequation}
\left(\Box-\mu^2 \right) \tilde{A}^{\mu}=0 \ .
\end{eqnarray}
This particle, referred to as the massive spin-1 dark graviton, is associated with a Yukawa potential that modifies the usual Newtonian gravity law. This modification leads to a model capable of explaining the dark sector's emergence, both in galaxies and at cosmological scales, as an apparent effect. In the following sections, we will explore the implications of this particle. Specifically, we begin by considering the coupling between the massive spin-1 gauge boson and the galaxy's baryonic matter fields in a spherically symmetric, static scenario with the four-potential $\tilde{A}_{\mu}$,
\begin{equation}
\tilde{A}_{\mu}=\left(\Phi(r),0,0,0\right) \ ,
\end{equation}
where $\Phi(r)$ is the gravitational potential due to the spin-1 graviton, then, we can write for the static case,
\begin{equation}
    \left(\nabla^2-\mu^2 \right) \Phi(r) = 0 \ .
\end{equation}
The solution is given in terms of the Yukawa potential $\Phi(r)=\Phi_{\text{YU}}$, where we have to set one of the constants to zero to avoid the divergent part (unphysical solution). Further, if we set $c_1=\pm \sqrt{\alpha_B G_N} M$\footnote{The choice of the gravitational charge $c_1$ is motivated from \cite{Jusufi:2024utf} and $\alpha_B $ denotes a dimensionless coupling parameter associated with the influence of the massive vector field.}, for the potential per unit mass, we get, 
\begin{equation}
\Phi_{\text{YU}}\left( r\right) = \sqrt{\alpha_B G_N} M \frac{e^{-\mu r}}{r} \ .
\end{equation}
The mass of the vector boson is related to the length scale by $\mu = 1/\lambda $. On galactic scales, we expect $ \lambda $ to be of the order of $kpc$. The total potential includes the modified Newtonian potential and the Yukawa potential. In the outer regions of galaxies, the repulsive force weakens due to the short range of the Yukawa potential, effectively explaining dark matter, as the dominant attractive corrections dominate due to the increase of Newton's constant, i.e., $G=G_N (1+\alpha)$.

%%%%%%%%%%%%%%%%%%%%%%%%%%%%%%%%%%%%%%%%%%%%%%%%%%%%%%%%%%%%%%%%%%%%%%%%%%%%%%%%%%%

\section{New black hole solution surrounded by massive vector fields}\label{Sec:Black hole solution surrounded by massive vector fields}

In what follows, we will find an exact black hole solution surrounded by a massive spin-1 graviton field. Let us choose the metric, 
\begin{equation}
    ds^2=-f(r)dt^2+\frac{dr^2}{g(r)}+r^2(d\theta^2+\sin^2\theta d\phi^2) \ ,
\end{equation}
along with the four potential, 
\begin{eqnarray}
   \tilde{A}_{\mu}=\left(\sqrt{\alpha_B G_N} M \frac{e^{- r/\lambda}}{r},0,0,0\right) \ .
\end{eqnarray}
In the strong-gravity regime, due to the coupling between a black hole and gauge fields, we anticipate a correction to the above energy-momentum tensor $T_{\mu \nu}^{\rm V} $. For a vanishing matter field $T_{\mu \nu}^{\rm M} =0$, near the black hole surrounded by massive vector fields, we expect the field equation to be, 
\begin{eqnarray}
 G_{\mu \nu}+\Lambda g_{\mu \nu} =8 \pi G\, \left(T_{\mu \nu}^{\rm V} +T_{\mu \nu}^{\rm int}(g_{\mu \nu}, \tilde{A}_{\mu})\right) \ ,
\end{eqnarray}
where $T_{\mu \nu}^{\rm int}(g_{\mu \nu}, \tilde{A}_{\mu})$ describes the interaction between the black hole and the massive vector field. But since we do not know the exact form of this interaction tensor, we shall  assume the following relation,
\begin{eqnarray}
    \mathcal{T}_{\mu \nu}=T_{\mu \nu}^{\rm V} +T_{\mu \nu}^{\rm int}(g_{\mu \nu}, \tilde{A}_{\mu}):=-\rho_V g_{\mu \nu}+t_{\mu \nu} \ ,
\end{eqnarray}
where $\rho_V$ is the energy density of the massive vector field, while the components of the tensor $t_{\mu \nu}$ will be determined. In addition, we shall simplify the calculations by supposing a vanishing cosmological constant. We assume the resulting energy-momentum tensor to have the components, 
\begin{eqnarray}
    {\mathcal{T}^{\mu}}_{\nu}=\left(-\rho_V,\, \mathcal{P}_r,\, \mathcal{P},\, \mathcal{P}\right) \ ,
\end{eqnarray}
with $\mathcal{P}=\mathcal{P}_{\theta}=\mathcal{P}_{\phi}$. Furthermore, we presuppose the conservation of the energy-momentum tensor to hold,
\begin{eqnarray}
    \nabla_\mu  \mathcal{T}^{\mu \nu}=0 \ .
\end{eqnarray}
Also, it is natural to choose $\rho_V=-\mathcal{P}_r$ and $f(r)=g(r)$. From the above equation, we get the following result,
\begin{eqnarray}\label{tmunu}
    t_{\mu \nu}={\rm diag} \left(0, 0, -(r/2)\, (\partial_r \rho_V), -(r/2)\, (\partial_r \rho_V)  \right) \ ,
\end{eqnarray}
meaning that 
\begin{eqnarray}
    \mathcal{P}=-\rho_V-(r/2)\, (\partial_r \rho_V) \ .
\end{eqnarray}

Using the energy-momentum tensor given by Eq. \eqref{emtensor}, we can obtain its components, 
\begin{eqnarray}
    {{T}^{\mu}}_{\nu}=\left(-\rho_V, P_r, P, P\right) \ ,
\end{eqnarray}
given by,
\begin{eqnarray}
    \rho_V=\frac{\alpha_B M^2 G_N e^{-\frac{2 r}{\lambda }} \left(f(r) (\lambda +r)^2+ \lambda ^2 \mu^2 r^2\right)}{8 \pi \lambda ^2 r^4 f(r)} \ ,
\end{eqnarray}
\begin{eqnarray}
    P_r=-\frac{\alpha_B M^2 G_N e^{-\frac{2 r}{\lambda }} \left(f(r) (\lambda +r)^2- \lambda ^2 \mu^2 r^2\right)}{8 \pi \lambda ^2 r^4 f(r)} \ ,
\end{eqnarray}
\begin{eqnarray}
    P=\frac{\alpha_B M^2 G_N e^{-\frac{2 r}{\lambda }} \left(f(r) (\lambda +r)^2+ \lambda ^2 \mu^2 r^2\right)}{8 \pi \lambda ^2 r^4 f(r)} \ .
\end{eqnarray}
At this point, we will define the following quantity
\begin{eqnarray}
    \bar{\mu}=\frac{\mu}{\sqrt{f(r)}} \ ,
\end{eqnarray}
which indeed describes the energy/mass of the particle measured at some large distance from the black hole. The energy-density simplifies as,
\begin{eqnarray}
    \rho_V=\frac{\alpha_B M^2 G_N e^{-\frac{2 r}{\lambda }} \left( (\lambda +r)^2+ \lambda ^2 \bar{\mu}^2 r^2\right)}{8 \pi \lambda ^2 r^4 } \ .
\end{eqnarray}
Using this result, we can determine the components of the stress-energy tensor \( t_{\mu\nu} \), which has a diagonal form as in Eq.~\eqref{tmunu}. Since the only non-zero components are $t_{\theta \theta}$ and $t_{\phi \phi}$ and they are equal, its value can be confirmed as, 

\begin{figure}[htb]
    \centering
    \includegraphics[scale=0.65]{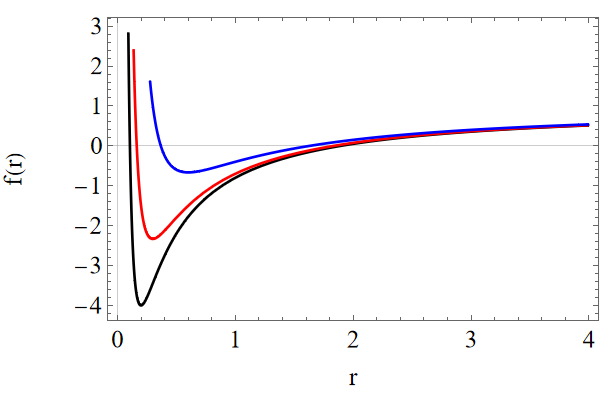}
    \caption{Plot of the metric function $f(r)$ for a fixed value for $\lambda=10^5$ and $\mathcal{M}=1$, using $\gamma=0.2$ (black curve), $\gamma=0.3$ (red curve), and $\gamma=0.6$ (blue curve), respectively.}
    \label{function}
\end{figure}

\begin{equation*}
     \frac{\alpha_B G_N M^2 e^{-\frac{2 r}{\lambda }} (\lambda +r) \left(2 \lambda ^2+r^2 \lambda ^2 \bar{\mu}^2+r^2 +2 \lambda  r\right)}{8 \pi  \lambda ^3 r^4} \ .
\end{equation*}
In the end, by using $\lambda = 1/\bar{\mu}$, we can compute the form of the components of the interacting energy-momentum tensor ${{T^{\rm int}}^\mu}_\nu = (-\rho^{\rm int},  P^{\rm int}_r, P^{\rm int}, P^{\rm int})$, turning out to be,
\begin{eqnarray}
    \left(0, \frac{\alpha_B G_N M^2 e^{-\frac{2 r}{\lambda }}}{4 \pi  \lambda ^2 r^2},  \frac{\alpha_B G_N M^2 e^{-\frac{2 r}{\lambda }}}{4 \pi  \lambda ^3 r} ,  \frac{\alpha_B G_N M^2 e^{-\frac{2 r}{\lambda }}}{4 \pi  \lambda ^3 r}\right) \ . \nonumber
\end{eqnarray}

Finally, from the Einstein field equations and using the choice $\lambda=1/\bar{\mu}$ along with $G=G_N (1+\alpha)$, we get the following exact solution, 
\begin{equation}
    f(r)=1-\frac{2 G_N \mathcal{M}}{r}+\frac{\alpha_B G_N^2 M^2 (1+\alpha) e^{-2 r/\lambda}}{r^2 \lambda}\left(r+\lambda \right) \ ,
\end{equation}
with 
\begin{eqnarray}
    \mathcal{M}=M(1+\alpha)
\end{eqnarray}
where $M$ is the black hole mass parameter, while $\mathcal{M}$ is the total mass of the black hole due to extra mass term $\alpha M$. That is an apparent type of mass. We can write the last equation further as, 
\begin{equation}\label{sol1}
 \boxed{   f(r)=1-\frac{2 G_N \mathcal{M}}{r}+\frac{\gamma\, G_N^2 \mathcal{M}^2 e^{-2 r/\lambda}}{r^2 \lambda}\left(r+\lambda \right)}
\end{equation}
where 
\begin{eqnarray}
    \gamma = \frac{\alpha_B}{1+\alpha} \ .
\end{eqnarray}
The effect of a massive vector field coupling to black hole spacetime essentially mimics the effect of an electric charge. In the special case where our parameter approaches the limit $\lambda \to \infty$, we recover the solution found in scalar-vector-tensor gravity \cite{Moffat:2014aja}.  
In Fig.~\ref{function}, we illustrate the presence of horizons for specific values of $\lambda$ and $\gamma$. 

In the following sections, we explore the phenomenological aspects of our solution, focusing on accretion disks, QNMs, stability, as well as the photon sphere and shadow radius.

%%%%%%%%%%%%%%%%%%%%%%%%%%%%%%%%%%%%%%%%%%%%%%%%%%%%%%%%%%%%%%%%%%%%%%%%%%%%%%
%\section{Accretion disk}\label{Sec:Accretion disk}

\section{Accretion disk}\label{Sec:Accretion disk}
The theoretical framework of geometrically thin accretion disks surrounding black holes \cite{Shakura:1972te,novikov1973astrophysics,page1974disk,thorne1974disk} posits that the accreting material is confined to a region near an axially symmetric plane. Quantitatively, this confinement is characterized by the condition $\cot\theta \ll 1$ within the physical coordinate system described by \eqref{sol1}. Furthermore, the disk material is typically assumed to move along nearly geodesic circular orbits in the equatorial plane ($\theta = \pi/2$), with the inner boundary of the disk located at the innermost stable circular orbit (ISCO). The radiation emitted by the accretion disk can be significantly influenced by the parameters of the black hole, making this study crucial for understanding the physical properties of black holes and their impact on astrophysical observations. In Refs. \cite{Jiang2024njc,Ditta2023vou,nozari2020quantum,Nozari:2024vxp,salahshoor2018circular} different black holes spacetime properties and their particle geodesic motions and accretions have been investigated extensively. In this section, we analyze the ISCO and generic circular orbits of massive particles in the gravitational field of the black hole described by \eqref{sol1}. Subsequently, we examine the radiative properties of the accretion disk, which play a fundamental role in determining its observational characteristics. 

\subsubsection{Circular orbits of massive particles}\label{CO}

To determine the ISCO radius, we analyze the equations of motion for a massive particle in the equatorial plane. A partial study of this problem was conducted in Ref. \cite{bakopoulos2024exact} using a different coordinate system. Here, we complete the analysis in the physical coordinate system, which, while algebraically more complex, provides greater physical clarity.

For orbits in the equatorial plane ($\theta = \pi/2$), the effective Hamiltonian of a massive particle is given by,

\begin{eqnarray}\label{Hamil}
{H}_{\eff}&=&\frac{1}{2}\left[g_{tt}\left(\frac{\di t}{\di\tau}\right)^2+g_{rr}\left(\frac{\di r}{\di\tau}\right)^2+g_{\phi\phi}\left(\frac{\di\phi}{\di\tau}\right)^2\right]\\\notag
&=&\frac{1}{2}\left[-f(r)\left(\frac{\di t}{\di\tau}\right)^2+\frac{1}{g(r)}\left(\frac{\di r}{\di\tau}\right)^2+r^2\left(\frac{\di\phi}{\di\tau}\right)^2\right],
\end{eqnarray}
where we have imposed the normalization condition on the four-velocity in the final step. Expressed in terms of the conserved specific energy and the angular momentum,
\begin{equation}\label{E}
E=-g_{tt}\frac{\di t}{\di\tau}=f(r)\frac{\di t}{\di\tau},~~~~L=g_{\phi\phi}\frac{\di\phi}{\di\tau}=r^2\frac{\di\phi}{\di\tau},
\end{equation}
Eq. \eqref{Hamil} is equivalent to the radial equation of motion,
\begin{equation}\label{reom}
\frac{1}{2}E^2=-\frac{1}{2}g_{tt}g_{rr}\left(\frac{\di r}{\di\tau}\right)^2+V_{\eff}(r)=\frac{1}{2}\left(\frac{\di r}{\di\tau}\right)^2+V_{\eff}(r).
\end{equation}
In this equation, the effective potential induced by the geometric configuration reads, 
\begin{equation}\label{Veff}
V_{\eff}(r)=-\frac{1}{2}g_{tt}\left(1+\frac{L^2}{g_{\phi\phi}}\right)=\frac{f(r)}{2}\left(1+\frac{L^2}{r^2}\right).
\end{equation}
The ISCO corresponds to a marginally stable circular trajectory. For such orbits, the conditions $( \frac{d r}{d \tau} = 0 )$ and $( \frac{d^2 r}{d \tau^2} = 0 )$ must hold. Additionally, the marginal stability condition requires $( \frac{d^3 r}{d \tau^3} = 0 )$. Applying these conditions to Eq.~\eqref{reom}, we obtain the following equivalent expressions 
\begin{equation}\label{VdVddV}
V_{\eff}(r)=\frac{1}{2}E^2,~~~~V'_{\eff}(r)=0,~~~~V''_{\eff}(r)=0.
\end{equation}
To determine the radius of the ISCO, we eliminate the parameter $L$ from the last two equations in \eqref{VdVddV} as,
\begin{equation}
\left[\frac{1}{g'_{tt}}\left(\frac{g_{tt}}{g_{\phi\phi}}\right)'\right]'=0.
\end{equation}
Thus, the radius of the ISCO surrounding the black hole \eqref{sol1} is determined by,
\begin{equation}
f(r)f''(r)-2f'(r)^2+\frac{3}{r}f(r)f'(r)=0.
\end{equation}
An analytical solution to this equation could not be obtained; therefore, we resorted to numerical methods for different values of parameters $\lambda$ and $\gamma$ as shown in Table \ref{t1}. For $\gamma=0$ the above equation simplifies to the Schwarzschild black hole case, where the ISCO radius is given by $r_{\rm isco} = 6M$. Table \ref{t1} demonstrates that for a fixed value of the $\lambda$ parameter, the increase of the $\gamma$ parameter leads to a reduction in the ISCO radius of a black hole immersed in massive vector fields within the framework of KK gravity.  Notably, in the context of Kaluza-Klein gravity with massive vector fields, the ISCO is generally situated at a greater radial distance from the event horizon than in the Schwarzschild and RN cases. Furthermore, for a fixed $\gamma$ parameter, an increase in the $\lambda$ parameter results in a further decrease in the ISCO radius of the black hole.\\

\begin{table*}
 \caption{Numerical values of $r_{\rm{isco}}$, maximum energy flux ($F_{\rm{max}}$), maximum temperature ($T_{\rm{max}}$), and efficiency ($\epsilon$) of an accretion disk of the black hole surrounded by massive vector fields in KK gravity (we set $M=1$).}
 \label{t1}
 \begin{tabular}{cccccc}
  \hline
  $\lambda$ &\hspace{5mm} $\gamma$ &\hspace{5mm} $r_{\rm{isco}}$ &\hspace{5mm} $F_{\rm{max}}$ ($\times10^{-6}$) & \hspace{5mm}$T_{\rm{max}}$ & \hspace{5mm}$\epsilon \%$ \\
  \hline
  & \hspace{5mm} 0.03 &\hspace{5mm} 8.934 & 1.952 & 0.0371 &\hspace{5mm} 5.758 \\
  & \hspace{5mm} 0.06 &\hspace{5mm} 8.867 & 1.993 & 0.0376 &\hspace{5mm} 5.799 \\
  100 &\hspace{5mm}  0.1 &\hspace{5mm} 8.777 & 2.066 & 0.0379 &\hspace{5mm} 5.855 \\
  &\hspace{5mm}  0.5 &\hspace{5mm} 7.789 & 2.205 & 0.038 &\hspace{5mm} 6.538 \\
  &\hspace{5mm}  0.8 &\hspace{5mm} 6.881 & 2.352 & 0.039 &\hspace{5mm} 7.301 \\
  \hline
  10 &\hspace{5mm}  &\hspace{5mm} 8.84 & 2.067 & 0.038 &\hspace{5mm} 5.818 \\
  100 &\hspace{5mm}  &\hspace{5mm} 8.777 & 1.144 & 0.032 &\hspace{5mm} 5.855 \\
  1000 &\hspace{5mm}  0.5 &\hspace{5mm} 8.771 & 0.576 & 0.027 &\hspace{5mm} 5.865 \\
  $10^{4}$ &\hspace{5mm}  &\hspace{5mm} 8.7709 & 0.472 & 0.025 &\hspace{5mm} 5.872 \\
  $10^{5}$ &\hspace{5mm}  &\hspace{5mm} 8.7708 & 0.408 & 0.024 & \hspace{5mm} 5.893 \\
  \hline
 \end{tabular}
\end{table*}

Beyond the ISCO, all circular orbits remain stable unless an outermost stable circular orbit exists. For general circular orbits only the first two equations of \eqref{VdVddV} are necessary. These equations can be utilized to determine the specific energy $E$ and the specific angular momentum $L$ of a test particle in a circular orbit of radius $r$, 
\begin{eqnarray}
\label{e}E&=&\frac{-g_{tt}}{\sqrt{-g_{tt}+g_{\phi\phi}\frac{g'_{tt}}{g'_{\phi\phi}}}}=\frac{f(r)}{\sqrt{f(r)-\frac{1}{2}rf'(r)}},\\
\label{l}L&=&\frac{g_{\phi\phi}\sqrt{-\frac{g'_{tt}}{g'_{\phi\phi}}}}{\sqrt{-g_{tt}+g_{\phi\phi}\frac{g'_{tt}}{g'_{\phi\phi}}}}=\frac{r\sqrt{\frac{1}{2}rf'(r)}}{\sqrt{f(r)-\frac{1}{2}rf'(r)}}.
\end{eqnarray}
The particle in this orbit has an angular velocity of 
\begin{equation}\label{Omega}
\Omega_{\varphi}=\frac{\di \phi}{\di t}=\frac{-g_{tt}L}{g_{\phi\phi}E}=\sqrt{-\frac{g'_{tt}}{g'_{\phi\phi}}}=\sqrt{\frac{f'(r)}{2r}},
\end{equation}
where we have used Eqs. \eqref{E}, \eqref{e}, and \eqref{l}. 

\begin{figure}
\centering
{\includegraphics[width=0.4\textwidth]{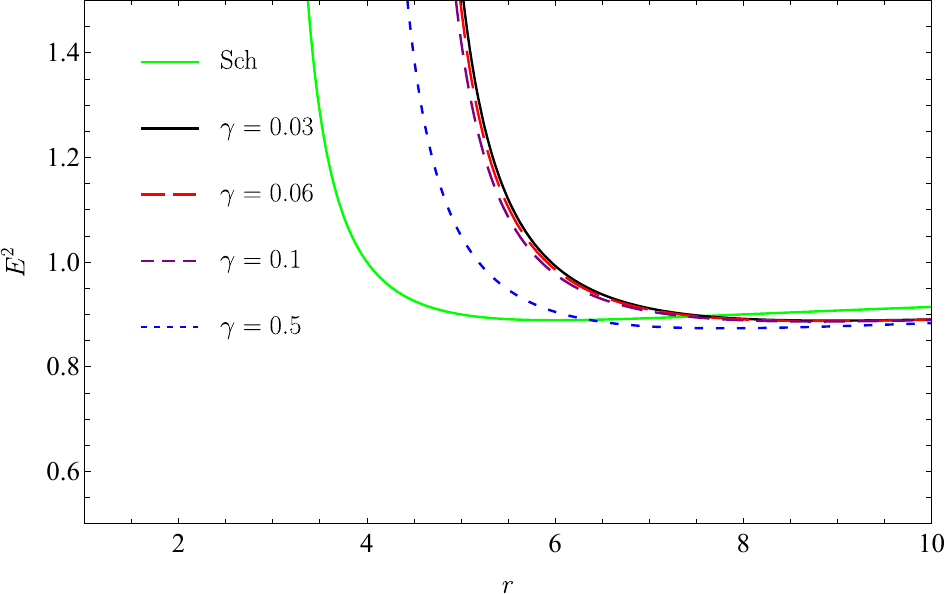}}
\,\,\,
{\includegraphics[width=0.4\textwidth]{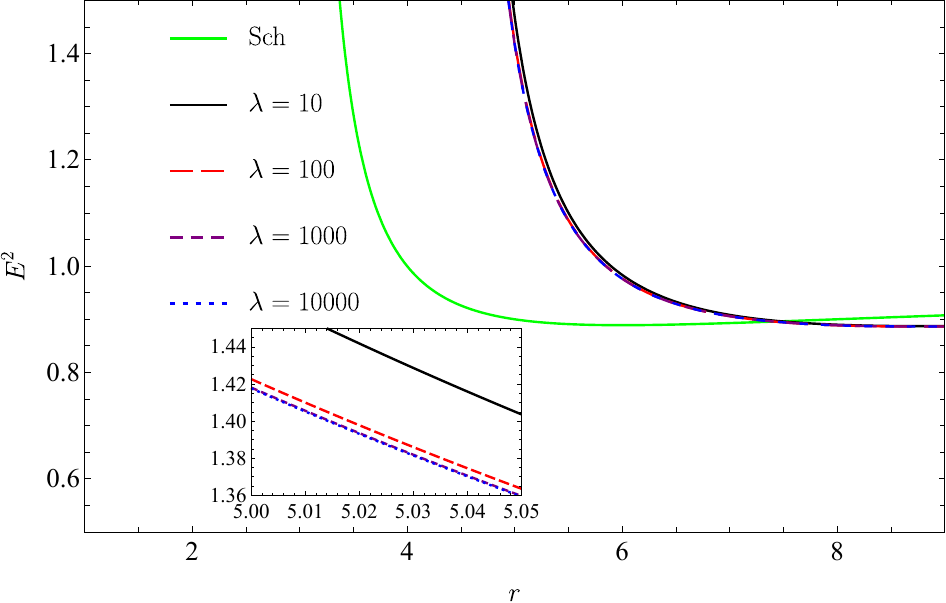}}
\caption{\label{FigE1} The specific energy of the black hole surrounded by massive vector fields in Kaluza-Klein gravity as a function of the radial coordinate $r$ for different values of $\gamma$ and $\lambda$.}
\end{figure}

\begin{figure}
\centering
{\includegraphics[width=0.4\textwidth]{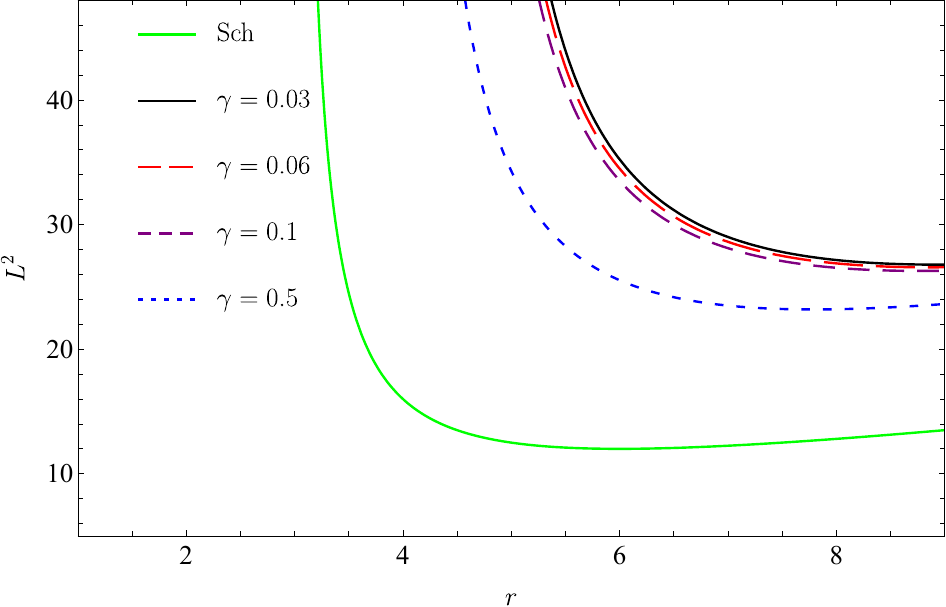}}
\,\,\,
{\includegraphics[width=0.4\textwidth]{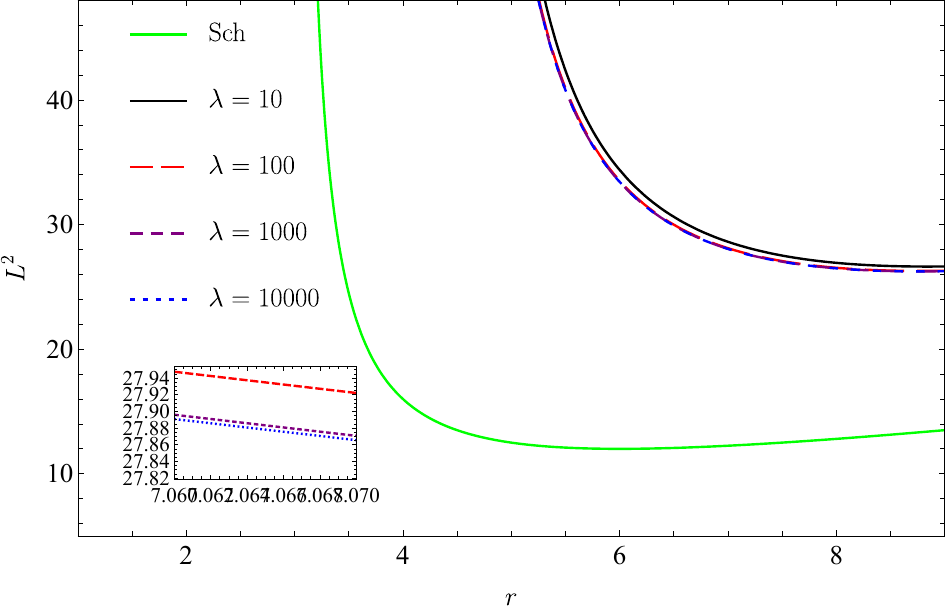}}
\caption{\label{FigL1} The specific angular momentum of the black hole surrounded by massive vector fields in Kaluza-Klein gravity as a function of the radial coordinate $r$ for different values of $\gamma$ and $\lambda$.}
\end{figure}

\begin{figure}
\centering
{\includegraphics[width=0.4\textwidth]{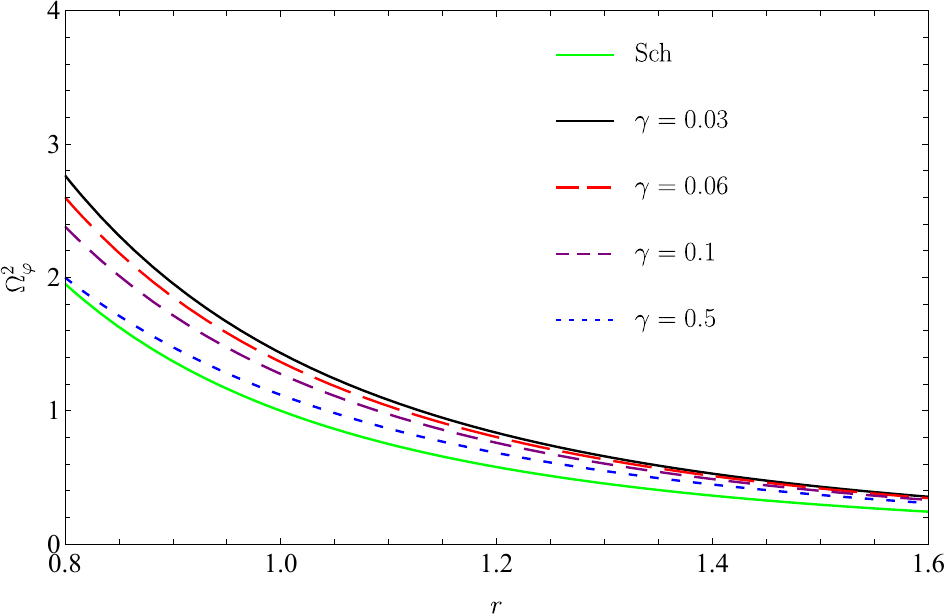}}
\,\,\,
{\includegraphics[width=0.4\textwidth]{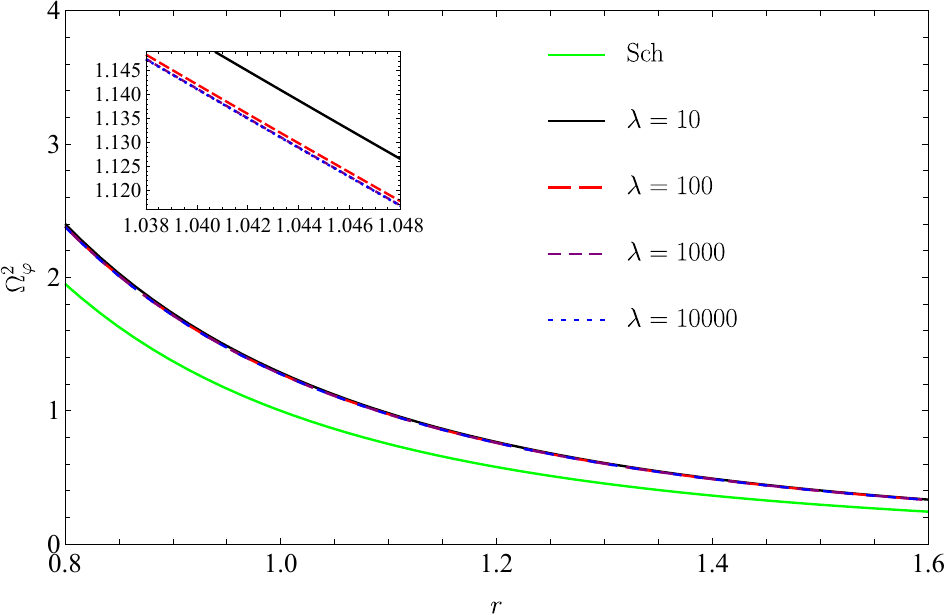}}
\caption{\label{FigO2} The angular velocity of the black hole surrounded by massive vector fields in Kaluza-Klein gravity as a function of the radial coordinate $r$ for different values of $\gamma$ and $\lambda$.}
\end{figure}

In the steady-state thin disk model, the specific energy $E$, specific angular momentum $L$, and angular velocity $\Omega_{\varphi}$ of a particle are determined by the radius of its orbit through Eqs.~\eqref{e}, \eqref{l}, and \eqref{Omega}, respectively, analogously to Kepler's third law. Setting $M = 1$, these quantities are depicted in the Figures~\ref{FigE1}, \ref{FigL1}, and \ref{FigO2} for different values of the $\lambda$ and $\gamma$ parameters, respectively. We can observe that by fixing the value of one parameter and raising the other, the above quantities increase in comparison to Schwarzschild black hole. At smaller radii, the influence of $\gamma$ and $\lambda$ becomes increasingly noticeable. Outside the ISCO, all of these quantities decrease with radius $r$. 

The specific energy $E$ at the ISCO is also of considerable importance. Assuming that all emitted photons escape to infinity, it can be used to determine the efficiency $\eta$ with which the black hole converts the accreting mass into radiation \cite{thorne1974disk},
\begin{equation}\label{eta}
\eta=\frac{\mathcal{L}_{\bol}}{\dot{M}c^2}\simeq1-E_{\isco},
\end{equation}

Here, $\mathcal{L}_{\bol}$ denotes the bolometric luminosity of the accretion disk, $\dot{M}$ represents the time-averaged mass accretion rate, and $E_{\isco}$ is the specific energy evaluated at the ISCO. Utilizing the data presented in Table \ref{t1}, we analyze the energy efficiency of the accretion disk around the black hole surrounded by massive vector fields in Kaluza-Klein gravity. The corresponding energy efficiency values are provided in Table \ref{t1}. The results indicate that the radiative efficiency increases with the parameter $\lambda$ and also due to the influence of the parameter $\gamma$. Remarkably, the Schwarzschild black hole case is recovered in the limit $\gamma=0$, where the efficiency is approximately $\eta \approx 5.72\%$ (see also \cite{kurmanov2022accretion}).

\subsubsection{Radiative properties of accretion disks}\label{CO2}

Assuming that each local region of an optically thick accretion disk is in thermodynamic equilibrium, the radiation emitted from the disk can be effectively described by black-body radiation with a position-dependent temperature. In this section, we will adopt the local black-body approximation in conjunction with the conservation equations to analyze the radiative properties of a geometrically thin and optically thick accretion disk surrounding a black hole, as described by the metric \eqref{sol1}.

The rest mass conservation equation connects the time-averaged rate of mass accretion onto the black hole, denoted by $\dot{M}$, to the surface density of the disk, $\Sigma(r)$ \cite{Feng:2024iqj},
\begin{equation}
\dot{M}=\frac{\di M}{\di t}=-2\pi\sqrt{-g_{tt}g_{rr}g_{\phi\phi}}\Sigma(r)\frac{\di r}{\di\tau}.
\end{equation}
If the whole material in the disk moves in a perfect circular orbit, i.e., $\frac{dr}{d\tau} = 0$, the mass accretion rate will be zero. In a disk with a nonzero accretion rate, particles gradually lose energy and angular momentum before accreting into the black hole. This mechanism converts the energy lost by particles into radiation. The energy flux emitted from the disk can be estimated using the equations of energy and angular momentum conservation \cite{Feng:2024iqj}, yielding the result \cite{Karimov:2018whx, Alloqulov:2024zln, Feng:2024iqj}, 
\begin{eqnarray}\label{flu}
\nonumber\mathcal{F}(r)&=&-\frac{\dot{M}}{4\pi\sqrt{-g_{tt}g_{rr}g_{\phi\phi}}}\frac{\Omega_{,r}}{\left(E-\Omega L\right)^2}\int_{r_{\rm{isco}}}^r\left(E-\Omega L\right)L_{,r}\di r\\
\nonumber&=&-\frac{\dot{M}\left(rf''-f'\right)}{4\pi r^2\sqrt{2rf'}\left(2f-rf'\right)}\\
&&\times\int_{r_{\rm{isco}}}^r\frac{\sqrt{r}\left(rff''-2rf'^2+3ff'\right)}{\sqrt{2f'}\left(2f-rf'\right)}\di r.
\end{eqnarray}

The Stefan-Boltzmann law, given by $\mathcal{F}(r) = \sigma_{SB} T^4(r)$, establishes a connection between the disk's radiation temperature, $T(r)$, and the energy flux, $\mathcal{F}(r)$, where $\sigma_{SB}$ represents the Stefan-Boltzmann constant. This relation implies that the radial variation of $T(r)$ is directly linked to the dependence of $\mathcal{F}(r)$ on $r$.

Additionally, the differential luminosity $\mathcal{L}_{\infty}$ observed at spatial infinity from a radius $r$ is connected to the flux, as described by \cite{novikov1973astrophysics,page1974disk,thorne1974disk},
\begin{eqnarray}\label{dL}
\nonumber\frac{\di\mathcal{L}_{\infty}}{d\ln r}&=&4\pi r\sqrt{-g_{tt}g_{rr}g_{\phi\phi}}E\mathcal{F}(r)\\
\nonumber&=&-\frac{\dot{M}f\left(rf''-f'\right)}{\sqrt{rf'}\left(2f-rf'\right)^{3/2}}\\
&&\times\int_{r_{\rm{isco}}}^r\frac{\sqrt{r}\left(rff''-2rf'^2+3ff'\right)}{\sqrt{2f'}\left(2f-rf'\right)}\di r.
\end{eqnarray}

\begin{figure}
\centering
\subfloat[\label{FigO1a} $\lambda=10$]{\includegraphics[width=0.4\textwidth]{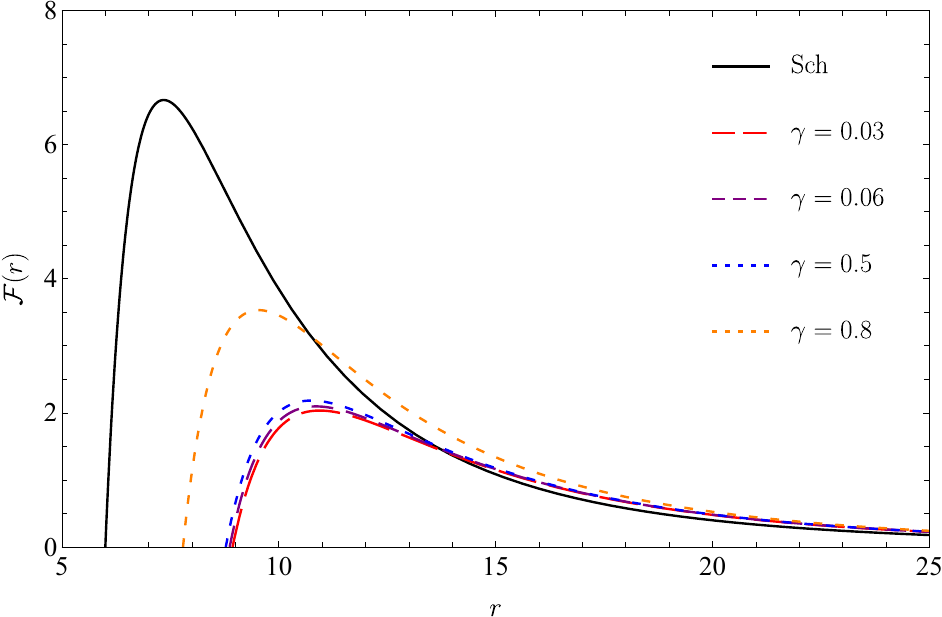}}
\,
\subfloat[\label{FigO1b} $\gamma=0.1$]{\includegraphics[width=0.4\textwidth]{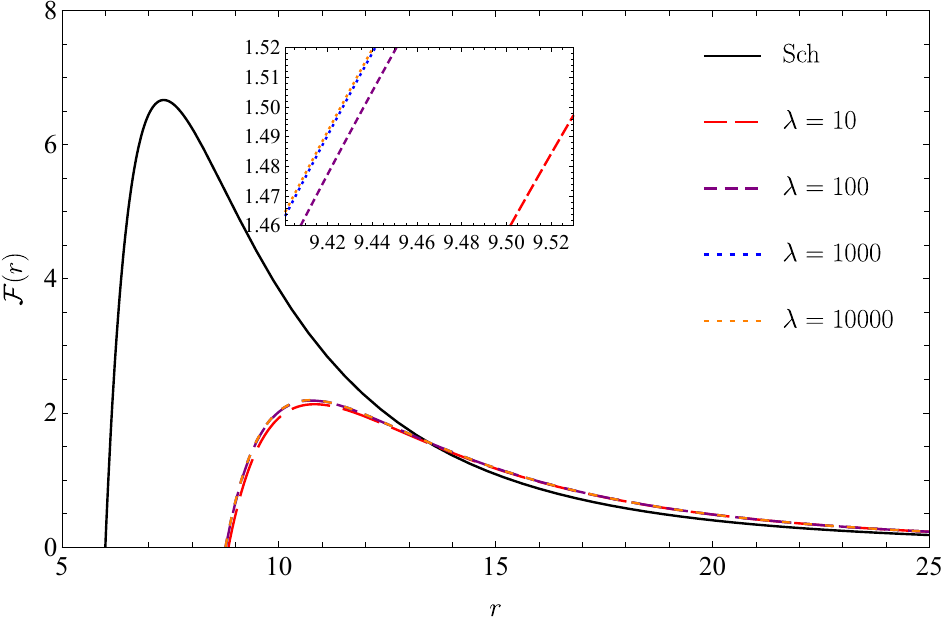}}
\caption{\label{FigO1}The energy flux (multiplied by $10^{-6}$) of an accretion disk around the black hole surrounded by massive vector fields in Kaluza-Klein gravity as a function of the radial coordinate $r$ for different values of $\gamma$ and $\lambda$.}
\end{figure}

\begin{figure}
\centering
\subfloat[\label{FigO2a} $\lambda=10$]{\includegraphics[width=0.4\textwidth]{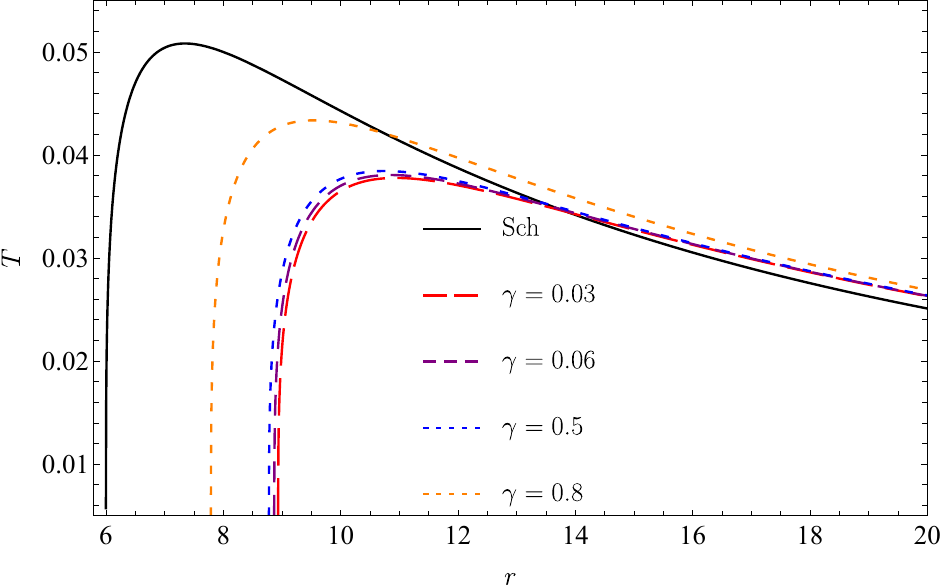}}
\,
\subfloat[\label{FigO2b} $\gamma=0.1$]{\includegraphics[width=0.4\textwidth]{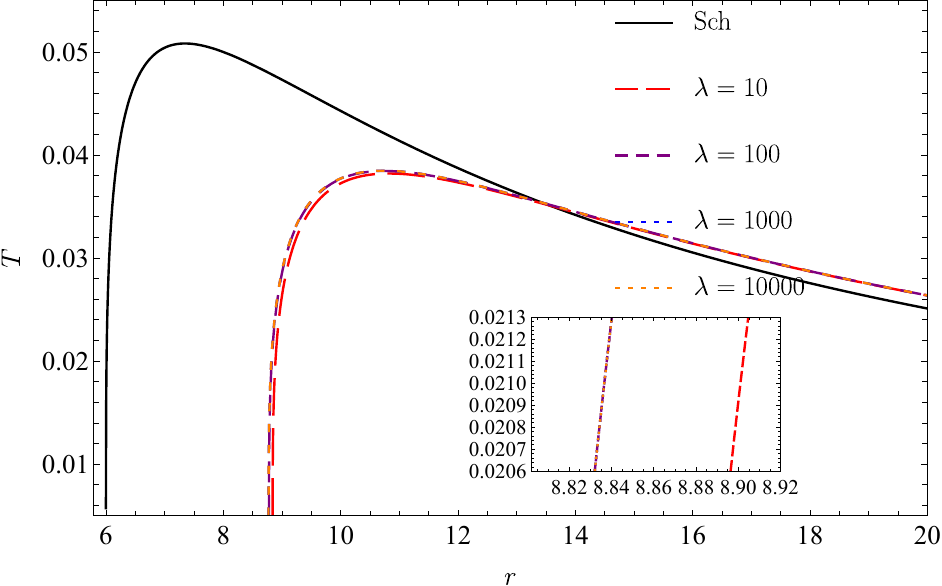}}
\caption{\label{FigT2}The radiation temperature $T(r)$ of an accretion disk around the black hole surrounded by massive vector fields in Kaluza-Klein gravity as a function of the radial coordinate $r$ for different values of $\gamma$ and $\lambda$.}
\end{figure}

\begin{figure}
\centering
\subfloat[\label{FigO3a} $\lambda=10$]
{\includegraphics[width=0.4\textwidth]{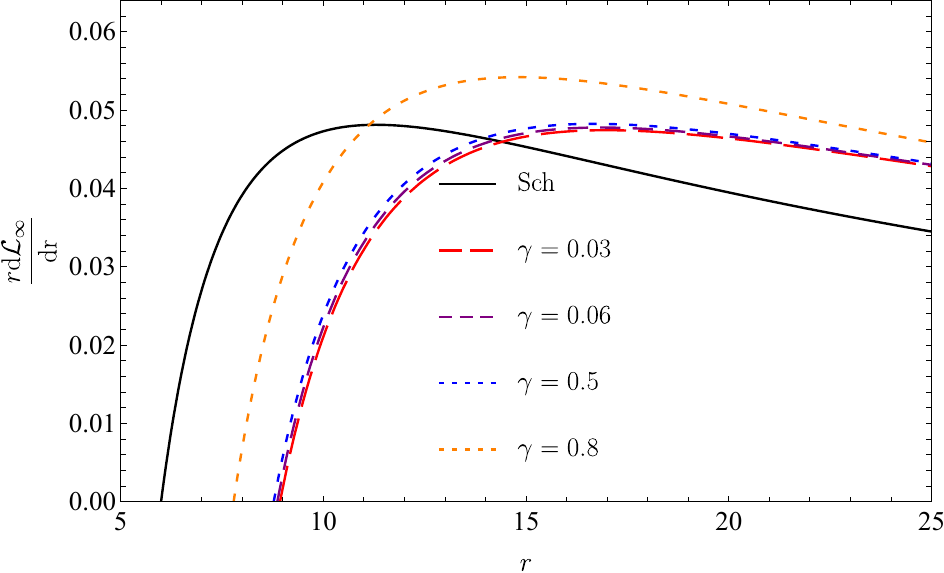}}
\,
\subfloat[\label{FigO3b} $\gamma=0.1$]{\includegraphics[width=0.4\textwidth]{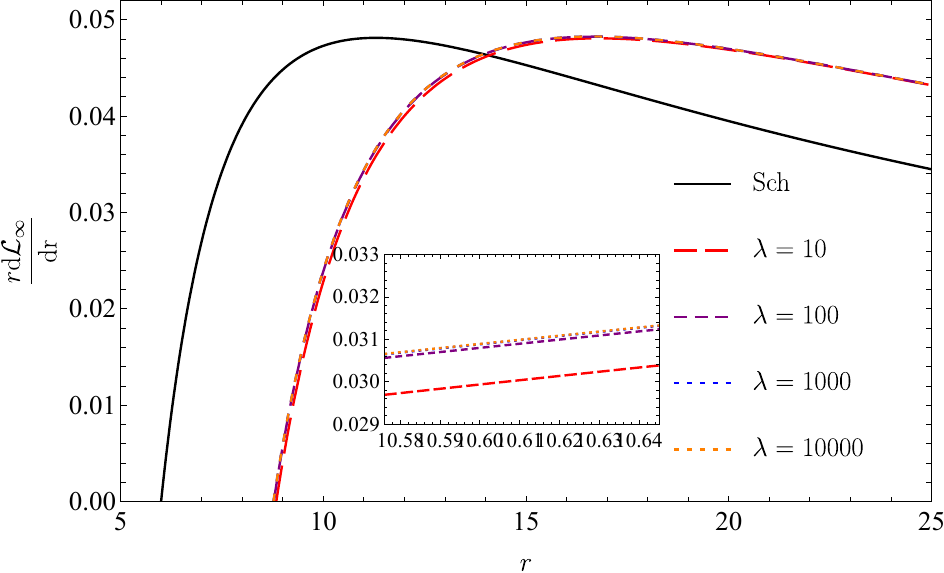}}
\caption{\label{FigO3}The differential luminosity  of an accretion disk around the black hole surrounded by massive vector fields in Kaluza-Klein gravity as a function of the radial coordinate $r$ for different values of $\gamma$ and $\lambda$.}
\end{figure}

Figures \ref{FigO1}, \ref{FigT2}, and \ref{FigO3} present numerical simulations of the energy flux \eqref{flu}, disk's temperature, and differential luminosity \eqref{dL}, respectively, which are normalized by $\dot{M}$. Furthermore, we systematically vary one parameter of the black hole spacetime surrounded by massive vector fields in KK gravity, while holding the other fixed in order to investigate its impact on these quantities. 

The energy flux distribution exhibits a characteristic pattern of an initial rise, followed by a peak, and then a gradual decline. Moreover, we observe that for a fixed parameter $\lambda$ and radial coordinate $r$, the radiation flux increases with increasing $\gamma$. Similarly, when $\gamma$ is held constant, an increase in $\lambda$ only results in a minimal enhancement of the radiation flux $\mathcal{F}(r)$. In fact, the effect of $\lambda$ on $\mathcal{F}(r)$ is quite negligible. In the distant region, all energy flux profiles appear to converge, remaining unaffected by the black hole parameters. Moreover, the black hole described in (\ref{sol1}) within KK gravity exhibits a lower energy flux than the Schwarzschild black hole for different values of $\lambda$ and $\gamma$. The temperature profile clearly demonstrates that as $\gamma$ increases, the temperature $T(r)$ also rises. Similarly, the lower plot, with $\gamma$ held constant, shows that an increase in the electric charge results in only a slight increase in $T(r)$. As illustrated in Fig \ref{FigO2}, the accretion disk surrounding the black hole defined in (\ref{sol1})  exhibits a lower temperature compared to the disks of Schwarzschild black holes for different values of $\lambda$ and $\gamma$. Also, the differential luminosity behavior closely matches the radiative flux depicted in Fig.~\ref{FigO1}.   

In the next section we will discuss the dynamical stability of the black hole solution \eqref{sol1} by using scalar perturbations and finding the response of the background geometry in the form of QNMs.

%%%%%%%%%%%%%%%%%%%%%%%%%%%%%%%%%%%%%%%%%%%%%%%%%%%%%%%%%%%%%%%%%%%%%%%%%%%%%%%%%%%%%%%%%%
\section{Quasinormal Modes and Stability}\label{Sec:Quasinormal Modes and Stability}

In order to check the stability of the new metric, we will consider a massive scalar perturbation $\Xi$,  which interacts with the background solution \eqref{sol1} and obeys the Klein-Gordon equation given by, 
\begin{equation}
    \frac{1}{\sqrt{-g}} \partial_\mu (\sqrt{-g} g^{\mu\nu} \partial_\nu \Xi ) - m^2 \Xi =0\,,
\end{equation}
where $m$ represents the mass of the scalar field. This equation can be decoupled from the angular part using the \textit{Ansatz},
\begin{equation}
    \Xi(t,r,\theta,\phi) =  e^{-i\omega t} \frac{X(r)}{r} Y_{\ell m}(\theta,\phi)\,,
\end{equation}
and accommodated in a Schr\"odinger-like form by means of a transformation to the so-called tortoise coordinate, which maps the space to $r_* \rightarrow \pm \infty$,
\begin{equation}
    r_* = \int \frac{dr}{f(r)}\,,
\end{equation}
resulting in the following expression,
\begin{equation}\label{schreq}
    \frac{d^2X}{dr_*^2} + Q(r_*) X = 0\,.
\end{equation}
Here $Q(r_*)=\omega^2 - V(r_*)$, $\omega=\omega_R + i\omega_I$ is the complex frequency associated to the mode, and the effective potential in terms of the radial coordinate reads,
\begin{equation}
    V(r) = \frac{f'f}{r} + \left[ \frac{\ell(\ell+1)}{r^2} + m^2\right] f \,.
\end{equation}

\begin{figure}[htb]
    \centering
    \includegraphics[scale=0.65]{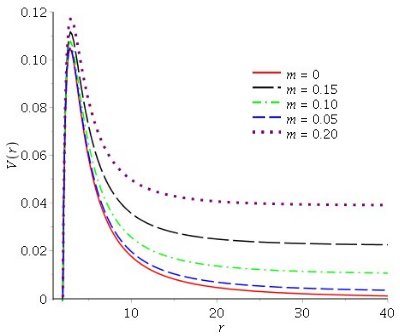}
    \caption{Effective potential for varying scalar perturbation masses and $\ell=1$.}
    \label{Vm}
\end{figure}

\begin{figure}[htb]
    \centering
    \includegraphics[scale=0.65]{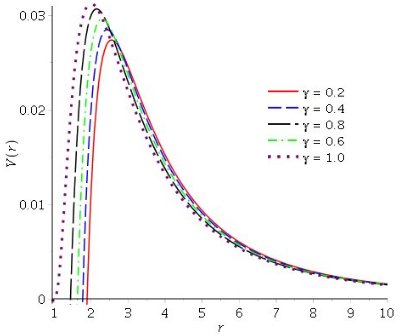}
    \caption{Effective potential for massless scalar perturbation with $\ell=0$ and varying $\gamma$.}
    \label{Vg}
\end{figure}

In Figs. \ref{Vm} and \ref{Vg} we show the shape of the effective potential varying the scalar field mass and the $\gamma$ parameter. After the event horizon, its general form displays a maximum and tends to a constant as the radial coordinate grows. The effect of the parameters can be seen in both figures. As $m$ grows, the peak is slightly enhanced, and the potential tends to $m^2$ for large $r$. Also, the more $\gamma$ increases, the higher the peak becomes. Comparing both graphs, we can also notice that the peak reaches higher values as the multipole number grows.

The solutions of Eq.\eqref{schreq} are the so-called quasinormal modes, ingoing plane waves crossing through the boundaries of $r_*$, which describe the response of the black hole when perturbed by test fields. Their complex frequency $\omega$ depends only on the black hole parameters, and a negative imaginary part signals a decaying wave, which guarantees the stability of the background geometry.  

In order to extract the quasinormal frequencies, we use a $6^{th}$-order WKB method~\cite{Konoplya2003}, whose motivation comes from the similarity between the equations of black hole perturbations and a 1-dimensional Schr\"odinger equation for a potential barrier. The method matches the exterior solutions, which are a linear combination of incoming and outgoing waves, very close to the peak,  simultaneously. In the internal region the function $Q(x)$ is Taylor-expanded around the peak and an asymptotic approximation to the interior solution is connected to the exterior ones. This procedure results in an algebraic equation for the quasinormal frequency,
\begin{equation}
    \frac{iQ_0}{\sqrt{2Q_2}}-\Lambda_2-\Lambda_3-\Lambda_4-\Lambda_5-\Lambda_6=n+\frac{1}{2}\,,
\end{equation}
where $Q_i$ represents the $i^{th}$-order derivative of $Q$ evaluated at the peak, $n \in \mathbb{N}$ labels the overtone number, and $\Lambda_i$ are coefficients depending on $Q_i$ given in~\cite{Iyer1987,Konoplya2003}. First-order WKB analysis indicates that, in general, increasing the value of the peak or the overtone number leads to a rise in the real $\omega_R$ and imaginary $\omega_I$ components of the QNF. Moreover, \(|\omega_R|\) exhibits a greater rate of increase compared to \(|\omega_I|\) \cite{Toshmatov:2015wga}.
Our results are shown in Tables \ref{tabAB} and \ref{tablm}. 

\begin{widetext}

\begin{table}[!h]
\arrayrulewidth=1pt
\renewcommand{\arraystretch}{1.5}
\begin{tabular}{c | c | c | c | c} 
\hline
$\alpha$ & $\alpha_B=0.2$ & $\alpha_B=0.5$ & $\alpha_B=0.7$ & $\alpha_B=1.0$ \\
\hline\hline
$\quad 0.5\quad $  &$\quad 0.11318-0.10146\,i\quad$ & $\quad 0.11781-0.10230\,i\quad $ & $\quad 0.12135-0.10282\,i\quad $ & $\quad 0.12843-0.10208\,i\quad $  \\
$\quad 1.0\quad$  & $\quad 0.11248-0.10131\,i\quad$ & $\quad 0.11579-0.10196\,i\quad$ & $\quad 0.11823-0.10237\,i\quad$ & $\quad 0.12333-0.10293\,i\quad $  \\
$\quad 1.5 \quad $  & $\quad 0.11206-0.10121\,i \quad $ & $\quad 0.11464-0.10175\, i \quad $ & $\quad 0.11650-0.10208\,i\quad $ & $\quad 0.11952-0.10257\,i\quad $ \\
$\quad 2.0 \quad $  & $\quad 0.11179-0.10115\,i\quad $ & $\quad 0.11390-0.10161\,i\quad $ & $\quad 0.11540-0.10189\,i\quad $ & $\quad 0.11781-0.10230\,i\quad $ \\
$\quad 2.5 \quad $  & $\quad 0.11160-0.10110\,i\quad $ & $\quad 0.11338-0.10150\,i\quad $ & $\quad 0.11464-0.10175\,i\quad $ & $\quad 0.11664-0.10211\,i\quad $ \\
$\quad 3.0 \quad $  & $\quad 0.11145-0.10107\,i\quad $ & $\quad 0.11300-0.10142\,i\quad $ & $\quad 0.11408-0.10164\,i\quad $ & $\quad 0.11579-0.10196\,i\quad $ \\ 
\hline 
\end{tabular}
\caption{\label{tabAB} Fundamental quasinormal frequencies ($\ell=0$) for varying $\alpha$ and $\alpha_B$. We considered $G_N=1$, $\mathcal{M}=1$, $\lambda = 10^5$, and $m=0$ (massless scalar perturbation).} 
\end{table} 

\begin{table}[!h]
\arrayrulewidth=1pt
\renewcommand{\arraystretch}{1.5}
\begin{tabular}{c | c | c | c | c} 
\hline
$\quad \ell\quad $ & $\quad m=0.05 \quad $ & $\quad m=0.10 \quad $ & $\quad m=0.15 \quad $ & $\quad m=0.20 \quad $ \\
\hline\hline
$\quad 1\quad $ & $\quad 0.30277-0.09791\,i\quad $ & $\quad 0.30599-0.09600\,i\quad $ & $\quad 0.31138-0.09277\,i\quad $ & $\quad 0.31896-0.08814\,i\quad $ \\
$\quad 5\quad $ & $\quad 1.09137-0.09712\,i\quad $ & $\quad 1.09247-0.09696\,i\quad $ & $\quad 1.09431-0.09668\,i\quad $ & $\quad 1.09689-0.09629\,i\quad $ \\
$\quad 10\quad $ & $\quad 2.08135-0.09708\,i\quad $ & $\quad 2.08194-0.09704\,i\quad $ & $\quad 2.08291-0.09696\,i\quad $ & $\quad 2.08428-0.09685\,i\quad $ \\
$\quad 50 \quad $ & $\quad 10.00651-0.09707\,i\quad $ & $\quad 10.00663-0.09707\,i\quad $ & $\quad 10.00684-0.09706\,i\quad $ & $\quad 10.00712-0.09706\,i\quad $ \\
$\quad 100 \quad $ & $\quad 19.91370-0.09707\,i\quad $ & $\quad 19.91375-0.09707\,i\quad $ & $\quad 19.91385-0.09707\,i\quad $ & $\quad 19.91400-0.09707\,i\quad $ \\
$\quad 1000\quad $ & $\quad 198.24440-0.09707\,i\quad $ & $\quad 198.24440-0.09707\,i\quad $ & $\quad 198.24441-0.09707\,i\quad $ & $\quad 198.24443-0.09707\,i\quad $ 
\\ 
\hline 
\end{tabular}
\caption{\label{tablm} Quasinormal frequencies for varying multipole number $\ell$ and perturbation mass $m$. We considered $G_N=1$, $\mathcal{M}=1$, $\lambda = 10^5$, $\alpha =2.0$, and $\alpha_B=0.5$.} 
\end{table} 

\end{widetext}

First of all, we did not found any instability in the model since all the frequencies have negative imaginary part. From Table \ref{tabAB} we notice that as $\alpha$ grows, the real part of the frequency decreases slowly as well as the imaginary part, although the latter always stays negative. The influence of $\alpha_B$ is the opposite, namely, the larger $\alpha_B$ becomes, the bigger the real frequency evolves. In the case of the imaginary frequency, as $\alpha_B$ increases, it slowly grows making the model even more stable. In terms of \(\gamma\), the real part of the quasinormal modes (QNMs) increases as the parameter \(\gamma\) increases.

In Table \ref{tablm} we observe that as the multipole number $\ell$ increases, the real frequency grows as expected and the imaginary frequency decreases, converging to a fixed value. In the eikonal limit (very large $\ell$) we notice that both the real and the imaginary parts converge to the same value independently of the mass of the perturbation. As a consequence, the effect of $m$ is only noticed for small values of the multipole number $\ell$. In that case, as $m$ increases, the real frequency grows while the imaginary part of the frequency decreases. The latter fact can be explained by looking at the potential in Fig. \ref{Vm}, where the peak height, measured from the asymptotic plateau level that corresponds to the perturbation mass squared value, gets shorter as $m$ increases, triggering less stability. In addition, we should stress that the eikonal limit of the real part of the frequency scales with the inverse of the shadow radius. This correspondence will be discussed in detail in the following section.

\section{Eikonal quasinormal modes and shadow radius correspondence}\label{Sec:QNMsand shadow}
Here, we aim to investigate the shadow of a black hole solution surrounded by a massive vector field. To achieve this goal, we begin with the Hamilton-Jacobi method for null geodesics in the black hole spacetime, expressed as  \cite{Perlick:2015vta},
\begin{equation}
\frac{\partial S}{\partial \sigma}+H=0,
\end{equation}
where $S$ is the Jacobi action and $\sigma$ is an affine parameter along the geodesics. If we consider a photon traveling along null geodesics in a spherically symmetric spacetime surrounded by matter, one can show that the Hamiltonian can be expressed as,  
\begin{equation}
\frac{1}{2}\left[-\frac{p_{t}^{2}}{f(r)}+f(r)p_{r}^{2}+\frac{p_{\phi}^{2}}{r^{2}}\right] =0.
\label{EqNHa}
\end{equation}
Due to the spacetime symmetries associated with the coordinates \( t \) and \( \phi \), there exist two constants of motion given by \( p_t = -E \) and \( p_{\phi} = L \), where \( E \) and \( L \) represent the energy and angular momentum of the photon, respectively. Next, the circular and unstable orbits correspond to the maximum value of the effective potential, satisfying the following conditions,  
\begin{equation}
V_{\rm eff}(r) \big \vert_{r=r_{\rm ph}}=0,  \qquad \frac{\partial V_{\rm eff}(r)}{\partial r}%
\Big\vert_{r=r_{\rm ph}}=0,  
\end{equation}
Without going into details, one can derive the following equation of motion,  
\begin{equation}
\frac{dr}{d\phi}=\pm r\sqrt{f(r)\left[\frac{r^{2}f(R)}{R^{2}f(r)} -1\right] }.
\end{equation}
Now, let us consider a light ray emitted by a static observer located at \( r_{0} \) and transmitted at an angle \( \vartheta \) with respect to the radial direction. Consequently, we have \cite{Perlick:2015vta},
\begin{equation}
\cot \vartheta =\frac{\sqrt{g_{rr}}}{\sqrt{g_{\phi\phi}}}\frac{dr}{d\phi}\Big\vert_{r=r_{0}}.
\label{Eqangle}
\end{equation}
Finally, the shadow radius of the black hole as observed by a static observer at \( r_0 \) is given by,  
\begin{equation}
R_{\rm sh}=r_{0}\sin \vartheta =R\sqrt{\frac{f(r_{0})}{f(R)}}\Bigg\vert_{R=r_{\rm ph}},
\end{equation}
where $r_{\rm ph}$ represents the photon sphere radius and $r_{0}$ is the position of a distant observer. It is worth noting that in an asymptotically flat spacetime, the observer's location at a sufficiently large distance implies $f(r_0) = 1$.

As is well known, QNMs are characterized by complex frequencies, $\omega=\omega_R+i \omega_I$, where the real part represents the oscillation frequency and the imaginary part is proportional to the damping of a given mode. Cardoso et al. \cite{Cardoso:2008bp} showed that, in the eikonal regime, the real part of QNMs is related to the angular velocity of the outermost photon orbit $\Omega_c$, while the imaginary part is proportional to the Lyapunov exponent $\tilde\lambda$, which determines the instability timescale of the orbit,
\begin{equation}
\omega_{\rm QNM}=\Omega_c \ell -i \left(n+\frac{1}{2}\right)|\tilde{\lambda}|\,.
\end{equation}
Additionally, Stefanov et al.~\cite{Stefanov:2010xz} established a connection between black hole eikonal QNMs and gravitational lensing observables, such as the flux ratio $\tilde r$ and the minimal impact angle $\theta_\infty$ in the strong-deflection regime. Later, a link between QNMs and the shadow radius of a black hole was further explored in~\cite{Jusufi:2019ltj} and analytically elaborated in~\cite{Cuadros-Melgar:2020kqn}. This correspondence indicates that the real part of the quasinormal frequency is inversely proportional to the shadow radius of the black hole.

To examine whether this correspondence holds for black holes modified by massive vector fields, we compute the QNM frequencies in the eikonal limit. Following~\cite{Cuadros-Melgar:2020kqn}, we consider the effective potential for scalar perturbations and employ the sixth-order WKB method. Expanding this expression for large $\ell$, i.e., taking the eikonal limit, we obtain a simple expression for the real part of the quasinormal frequency,
\begin{equation}\label{omre1}
\omega_R = \left(l+\frac{1}{2}\right) \frac{\sqrt{f(r_{\rm ph})}}{r_{\rm ph}}\,.
\end{equation}
To determine the peak of the potential, we note that in the eikonal limit the effective potential has the same form as the lightlike geodesic potential. Thus, the maximum of the scalar perturbation potential corresponds to the photon sphere radius, i.e., $r_{\rm ph} = r_c$. Consequently, for the shadow radius $R_{\rm sh}$, we can write  
\begin{equation}\label{omre2}
\omega_R = \left(\ell+\frac{1}{2}\right) \frac{1}{R_{\rm sh}}\,.
\end{equation}
Thus, the correspondence still holds, but an additional factor depends on the radial coordinate of the distant observer $r_0$. Furthermore, for real astrophysical values of Sgr A$^*$, as discussed in the previous section, $f(r_0)$ approaches unity, leading to the standard result found in~\cite{Cuadros-Melgar:2020kqn,Jusufi:2019ltj}.

\begin{figure}[htb]
    \centering
    \includegraphics[scale=0.6]{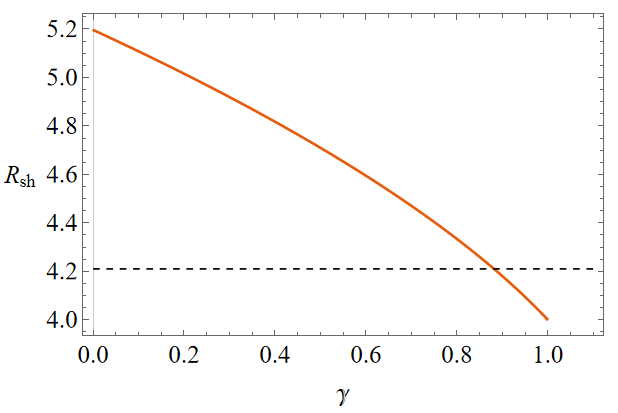}
    \caption{The shadow radius by varying $\gamma$. We fixed $\mathcal{M}=1$ and $\lambda=10^5$.}
    \label{shradius}
\end{figure}

\begin{figure}[htb]
    \centering
    \includegraphics[scale=0.6]{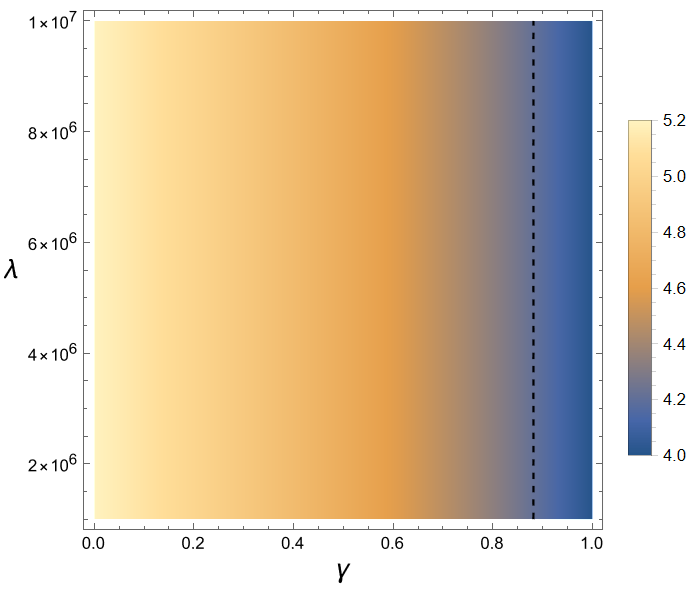}
    \caption{A density plot of the shadow radius by varying $\gamma$ and $\lambda$. }
    \label{shradius2}
\end{figure}

\begin{figure}[htb]
    \centering
    \includegraphics[scale=0.6]{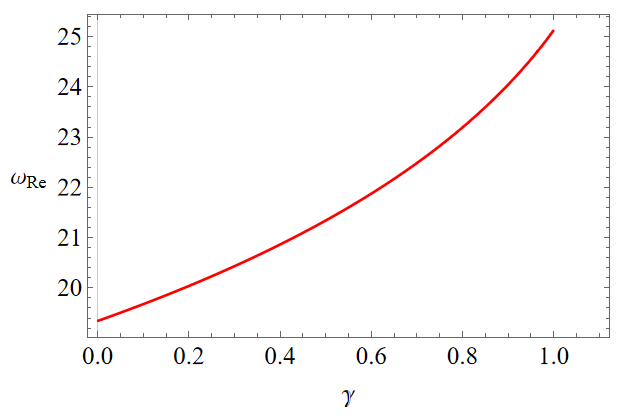}
    \caption{The real part of QNMs by varying $\gamma$. We fixed $\mathcal{M}=1$ and $\lambda=10^5$ and $l=100$.}
    \label{omegar1}
\end{figure}

\begin{figure}[htb]
    \centering
    \includegraphics[scale=0.6]{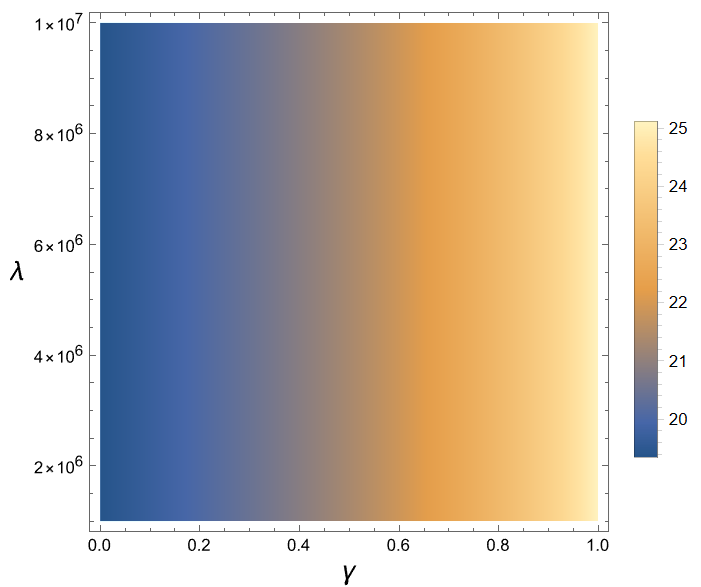}
    \caption{A density plot of the real part of QNMs by varying $\gamma$ and $\lambda$ and $\ell=100$.}
    \label{omegar2}
\end{figure}

Let us now proceed to constrain the black hole parameters using EHT observations of Sgr A*. In order to achieve this, we first fix $\mathcal{M} = 1$, which is crucial for constraining the dark-matter parameter $\gamma$. Using EHT observations of Sgr A*, it was shown that within $2\sigma$ constraints \cite{Vagnozzi:2022moj},  
\begin{eqnarray}
4.21 \lesssim R_{\rm sh}/\mathcal{M} \lesssim 5.56\,,
\end{eqnarray}  
which yields the upper bound $\gamma \lesssim 0.88$ (within $2\sigma$). This result is illustrated in Fig. \ref{shradius}, where we also observe that the presence of the surrounding vector field reduces the shadow radius. This effect is analogous to that of an electric charge (see \cite{Vagnozzi:2022moj}).  

Although $\gamma$ can be constrained, Fig. \ref{shradius2} indicates that there is a degeneracy, preventing us from placing constraints on $\lambda$. On the other hand, from Figs. \ref{omegar1} and \ref{omegar2} we observe that the real part of the quasinormal modes (QNMs) increases as $\gamma$ increases. These results are consistent with QNM values obtained via the WKB approximation in Sec.\ref{Sec:Quasinormal Modes and Stability}, confirming that the correspondence between QNMs and the shadow radius holds.

%%%%%%%%%%%%%%%%%%%%%%%%%%%%%%%%%%%%%%%%%%%%%%%%%%%%%%%%%%%%%%%%%%%%%%%%%%%%%%
\section{Conclusions}\label{Sec:Conclusions}

Our study presents an exact black hole solution surrounded by a massive vector field in the Kaluza-Klein gravity framework. Through dimensional reduction, five-dimensional KK gravity is reformulated in four dimensions with a scalar field $\Phi $ and gauge fields  $A_\mu$. Using ideas from superconductivity, one can introduce an interaction between the gauge and complex scalar field that induces a mass term for spin-1 gravitons, modifying gravity at long ranges via Yukawa-type corrections. These deviations from the inverse-square law have potential implications for extended gravity theories and astrophysical observations. Since the black hole is immersed in a surrounding vector field, its interaction with the vector field introduces an additional interaction term in the Einstein field equations. We have explicitly derived the form of this interaction term and, by imposing a spherically symmetric \textit{Ansatz}, we have solved the modified Einstein equations to obtain an exact black hole solution. The presence of the vector field significantly alters the spacetime structure, leading to deviations from classical general relativity solutions. 

We then studied the effect of the vector field on the accretion disk matter and QNMs. We first investigated the motion of neutral particles, circular orbits, and the radiation properties of the accretion disk surrounding a black hole in the presence of massive vector fields within the framework of Kaluza-Klein gravity. It is observed that by fixing the value of one parameter and increasing the other, the values of the energy $E^{2}$, the angular momentum $L^{2}$, and the frequency $\Omega_{\varphi}^{2}$ increase when compared to the Schwarzschild black hole. Furthermore, at smaller radii, the influence of the parameters $\gamma$ and $\lambda$ becomes more pronounced. Our findings indicate that increasing the vector field parameter $\lambda$ as well as the parameter $\gamma$ leads to a reduction in the ISCO radius of test particles. Furthermore, we investigated the radiative efficiency, and the results are summarized in Table \ref{t1}. It is observed that the radiative efficiency increases as the parameters $\lambda$ and $\gamma$ grow. Finally, we considered the accretion disk around a black hole surrounded by massive vector fields as a primary source of information regarding the surrounding spacetime geometry and its nature in Kaluza-Klein gravity. To provide valuable insights into the unique properties of the black hole in this framework, we examined the influence of the parameters $\lambda$ and $\gamma$ on the radiative properties of the accretion disk, including the flux of electromagnetic radiation, the temperature, and the differential luminosity of the disk. Interestingly, we found that the curves of these accretion disk radiation quantities shift upwards toward higher values as the parameters $\gamma$ and $\lambda$ increase, although only a slight increase was observed in the case of rising $\lambda$. This led to a decrease in these quantities when compared to the Schwarzschild black hole in Einstein gravity.

When analyzing the QNMs of massive scalar perturbations, we found no instabilities. Thus, the black hole surrounded by massive vector fields is stable, although we should stress that metric perturbations should give a definitive answer on these grounds. Regarding the influence of the black hole parameters, a growing $\alpha$ makes the real and imaginary frequencies decrease slowly. Meanwhile, as $\alpha_B$ increases, the real and imaginary parts of the frequency also grow. The effect of the scalar field mass is only noticed for small multipole numbers. As $\ell$ grows, the frequencies are $m$-independent and converge to a fixed value, which signals the eikonal limit. Moreover, when studying the correspondence of the QNMs with the shadows, we noticed that the effect of the coupling is analogous to that of an electric field. In particular, the real part of the QNMs increases with the increase in the coupling between the black hole and the massive vector field. On the other hand, the shadow radius decreases as this parameter increases. This demonstrates that in our solution, the correspondence between QNMs and the shadow radius holds. Furthermore, using EHT observations of Sgr A*, we found an upper bound of $\gamma \lesssim 0.88$ (within $2\sigma$).  

\section*{Acknowledgments}
A.A. is financially supported by the institute post-doctoral fellowship of IITK. B.C-M. thanks fruitful discussions with A.B. Pavan and C.E. Pellicer.

\bibliography{main}

\end{document}